\documentclass{aa}
\usepackage{graphicx}
\usepackage{txfonts}
\usepackage{natbib}
\def\figpath{}

\def \nptt {111}
\def \npte {69}
\def \Vrad {-31.666}

\def \wrms {11.87} 
\def \chs  {1.494}

\def \per  {3.69}
\def \perd {1349}

\def \Tpr  {51236}
\def \ecc  {0.462}
\def \omg  {91.4}
\def \Kap  {40.48}

\def \asn  {4.45}
\def \fm   {6.5 \times 10^{-6}}
\def \mss  {2.96}
\def \Mss  {0.0028}
\def \sma  {2.63}

\def \perB {55.43}

\def \TprB {45928}
\def \eccB {0.409}
\def \omgB {227.6}
\def \KapB {2041}

\def \asnB {3466}
\def \fmB  {13.56}
\def \mssB {351.5}
\def \MssB {0.34}
\def \smaB {17.23}
\def \perq {1347}  
\def \eccq {0.44}  
\def \mssq {2.9}  
\def \rmsq {14.61} 
\def \chiq {1.811}  

\begin{document}
   \title{The ELODIE survey for northern extra-solar planets}
   \subtitle{IV. {HD}\,196885, a close binary star with a 3.7-year planet}

   \titlerunning{{\small HD}\,196885, a close binary star with a 3.7-year planet}


   \author{A.C.M. Correia\inst{1,2} \and S. Udry\inst{2} \and M. Mayor\inst{2}
           \and A. Eggenberger\inst{3,2} \and D. Naef\inst{4} \and J.-L.
	   Beuzit\inst{3} \and C. Perrier\inst{3} \and \\ D. Queloz\inst{2} \and
	   J.-P. Sivan\inst{5} \and F. Pepe\inst{5} \and N.C. Santos\inst{6,2}
	   \and D. S\'egransan\inst{5}}

   \authorrunning{A.C.M. Correia et al. }

   \offprints{A.C.M. Correia, \email{correia@ua.pt}}

\institute{Departamento de F\'\i sica da Universidade de Aveiro, Campus
           Universit\'ario de Santiago, 3810-193 Aveiro, Portugal
   \and
           Observatoire de Gen\`eve, 51 Ch. des Maillettes, 1290 Sauverny,
           Switzerland
   \and
           Laboratoire d'Astrophysique de Grenoble, Universit\'e Joseph Fourier,
	   BP 53, 38041 Grenoble Cedex 9, France 
   \and
           European Southern Observatory, Casilla 19001, Santiago 19,
           Chile
   \and
           Laboratoire d'Astrophysique de Marseille, Traverse du Siphon
	   BP 8, 13376 Marseille Cedex 12, France
   \and
           Centro de Astrof\'isica da Universidade do Porto, Rua das Estrelas,
	   4150-762 Porto, Portugal	      
	      }

   \date{Received 01 October, 2006; accepted ?? }


  \abstract
   {}
   {We aim significantly increase the number
   of detected extra-solar planets in a magnitude-limited sample to improve our
   knowledge of their orbital elements distributions and thus obtain better
   constraints for planet-formation models.} 
   {Radial-velocity data were taken at Haute-Provence Observatory (OHP, France)
   with the ELODIE echelle spectrograph.}
   {We report the presence of a planet orbiting {\small HD}\,196885\,A,
   with an orbital period of \perd~days. This star was previously suggested to host a
   $ 386 $-day planet, but we cannot confirm its existence. We also
   detect the presence of a stellar companion, {\small HD}\,196885\,B, and give
   some constraints on its orbit. } 
   {}


   \maketitle
%


\section{Introduction}

The {\it {\small ELODIE} Planet Search Survey} was an extensive radial-velocity
northern survey of dwarf stars at the Haute-Provence Observatory (OHP,
France) using the {\small ELODIE} high-precision fiber-fed echelle spectrograph 
\citep{Baranne_etal_1996} mounted at the Cassegrain focus of the 1.93-m
telescope. 
It started to operate at the end of 1993 and acquired data until the summer of
2006, when it was replaced by the new echelle spectrograph {\small SOPHIE}
\citep{Bouchy_etal_2006}.

This survey is part of a large effort aimed at an extra-solar
planet search through radial-velocity measurements, in order to characterize the
types of exoplanets and to bring strong constraints on their processes of
formation and evolution.
{\small ELODIE} was responsible for finding many extra-solar
planets, among them the first hot Jupiter, 51\,Peg\,b \citep{Mayor_Queloz_1995}. 
The original sample consisted of 142 stars, but a new sample of 330 stars was
defined in 1997. Details of the program and the surveyed sample can be
found in \citet{Perrier_etal_2003}. 

In this paper we present the detection of two bodies around the
star {\small HD}\,196885\,A, one in the mass range of planets and the other
believed to be a stellar companion.
This star was previously described to harbor a planet with a period of about
386~days (although no published reference is known),
but we could not confirm its presence in our
observations (our analysis shows a planet with an orbital period of about
\perd~days instead).
The stellar companion was also recently observed using NACO adaptative optics by
\citet{Chauvin_etal_2006,Chauvin_etal_2007}.
The stellar properties of {\small HD}\,196885\,A are briefly recalled in
Sect.\,2 and the radial velocities with the two detected companions are
described in Sect.\,3. In Sect.\,4 we discuss the possibility of the existence
of more companions, in particular with orbital periods around $ 368 $~days.
Last section is devoted to our conclusions.


\section{HD\,196885\,A stellar characteristics}

{\small HD}\,196885\,A was observed by the {\small HIPPARCOS}
astrometric satellite ({\small HIP}\,101966).  
A high-precision spectroscopic study of this star was also performed by
\citet{Sousa_etal_2006} in order to examine the metallicity distribution of 
stars hosting planets.
Tßhis star was also studied by near-infrared survey with adaptive
optics of faint circumstellar environments, sensitive to companions within the
stellar and the sub-stellar domains \citep{Chauvin_etal_2006,Chauvin_etal_2007}.
Observed and inferred stellar parameters from 
these different sources are summarized in Table\,\ref{T1}.

In the {\small HIPPARCOS} catalogue, {\small HD}\,196885\,A is given a spectral
type F8~IV, a visual magnitude $ V = 6.398 $ and a color index $ B - V = 0.559 $. 
The measured parallax ($ 30.31 \pm 0.81 $ mas) leads to a distance of $ 33.0 \pm
0.9 $~pc and an absolute magnitude of $ M_V = 3.8 \pm 0.1 $.
From {\small CORALIE} spectra, \citet{Sousa_etal_2006} derived an effective
temperature $ T_\mathsf{eff} = 6340 \pm 39 $~K, a gravity log $ g = 4.46 \pm
0.02 $ and a high metal content [Fe/H] $ = 0.29 \pm 0.05 $ (Table\,\ref{T1}).
The bolometric correction ($ BC = -0.006 $) is computed from \citet{Flower_1996}
using the spectroscopic $T_\mathsf{eff}$ determination.
The bolometric magnitude is then $ M_{Bol} = 3.791 $, which allows us to 
derive a stellar luminosity of $ L = 2.40\,L_\odot $.
\citet{Sousa_etal_2006} also computed the mass, $ M = 1.33\,M_\odot$, 
from evolutionary tracks using Geneva models 
\citep{Schaller_etal_1992, Schaerer_etal_1993}.
From the $ B-V $ value and {\small ELODIE} correlation functions we find $ v
\sin i = 7.3 \pm 1.5 $~[km/s], meaning that the star is rotating fast.
However, its measured activity level is very weak: $ \log R'_{HK} =
-5.01 $ \citep{Wright_etal_2004}. From this value, these authors inferred a rotation period of about 15~days.
According to \citet{Pace_Pasquini_2004}, the activity level becomes weak 
and constant after $\sim$1.5~Gyr, setting a lower limit on the age of the star.
On the other hand,
following the Bayesian approach described in \citet{Pont_Eyer_2004} we
estimate a maximum age of 2.5~Gy. 
The derived stellar atmospheric parameters and age
are compatible with the expected values for a high metallicity
late-F dwarf (F8~V), and are slightly at odds with the evolutionary
status given by the {\small HIPPARCOS} catalogue.

\begin{table}[t!]
\caption{Observed and inferred stellar parameters for {\small HD}\,196885\,A.}    
\label{T1} 
\begin{center}
\begin{tabular}{l l c} \hline \hline
\noalign{\smallskip}
{\bf Parameter}  & $ \quad \quad \quad \quad \quad \quad $ 
& {\bf {\small HD}\,196885\,A}\\
\hline 
\noalign{\smallskip}
Spectral Type & & F8\,V \\ 
$ V   $ &       & $ 6.398 $  \\ 
$ B-V $ &       & $ 0.559 \pm 0.006 $  \\ 
$ \pi $ & [mas] & $ 30.31 \pm 0.81 $  \\ 
$ d   $ & [pc]  & $ 33.0  \pm 0.9 $  \\ 
$ M_V $ &       & $ 3.8   \pm 0.1 $  \\ 
$ BC  $ &       & $ - 0.006 $  \\ 
$ M_{Bol} $ &   & $ 3.79 $  \\ 
$ L  $  & $ [L_\odot] $ & $ 2.40 $ \\ 
\hline
\noalign{\smallskip}
 [Fe/H] &       & $ 0.29 \pm 0.05 $  \\ 
$ \log R'_{HK} $ & &      $-5.01 $ \\ 
$ P_\mathrm{rot.} $   &  [days]  & $ 15. $  \\ 
$ M $   & $ [M_\odot] $ & $ 1.33 $  \\ 
$ T_\mathrm{eff} $ & [K] & $ 6340 \pm 39 $  \\
 log $ g $ & [cgs] & $ 4.46 \pm 0.02 $  \\
$ v \sin i $ & [km/s] & $ 7.3 \pm 1.5 $  \\
 age & [Gyr] & $ 2.0 \pm 0.5 $   \\ 
\hline
\end{tabular}
\end{center}
Photometric and spectral type are from \citet{Chauvin_etal_2006}.
Astrometric parameters are from {\small HIPPARCOS} \citep{ESA_1997}. 
The mass $ M $ and the atmospheric parameters $T_\mathsf{eff}$, log $g$ and
[Fe/H] are from \citet{Sousa_etal_2006}.  
The bolometric correction is computed from \citet{Flower_1996} using the
spectroscopic $T_\mathsf{eff}$ determination.
The activity level and the rotation period are from \citet{Wright_etal_2004}.
The given age was obtained following the Bayesian approach of
\citet{Pont_Eyer_2004}. 
\end{table}




\section{Orbital solutions for the HD\,196885 system}
\label{orbsolutions}

\begin{figure}[t!]
   \centering
    \includegraphics*[height=8.5cm,angle=270]{\figpath 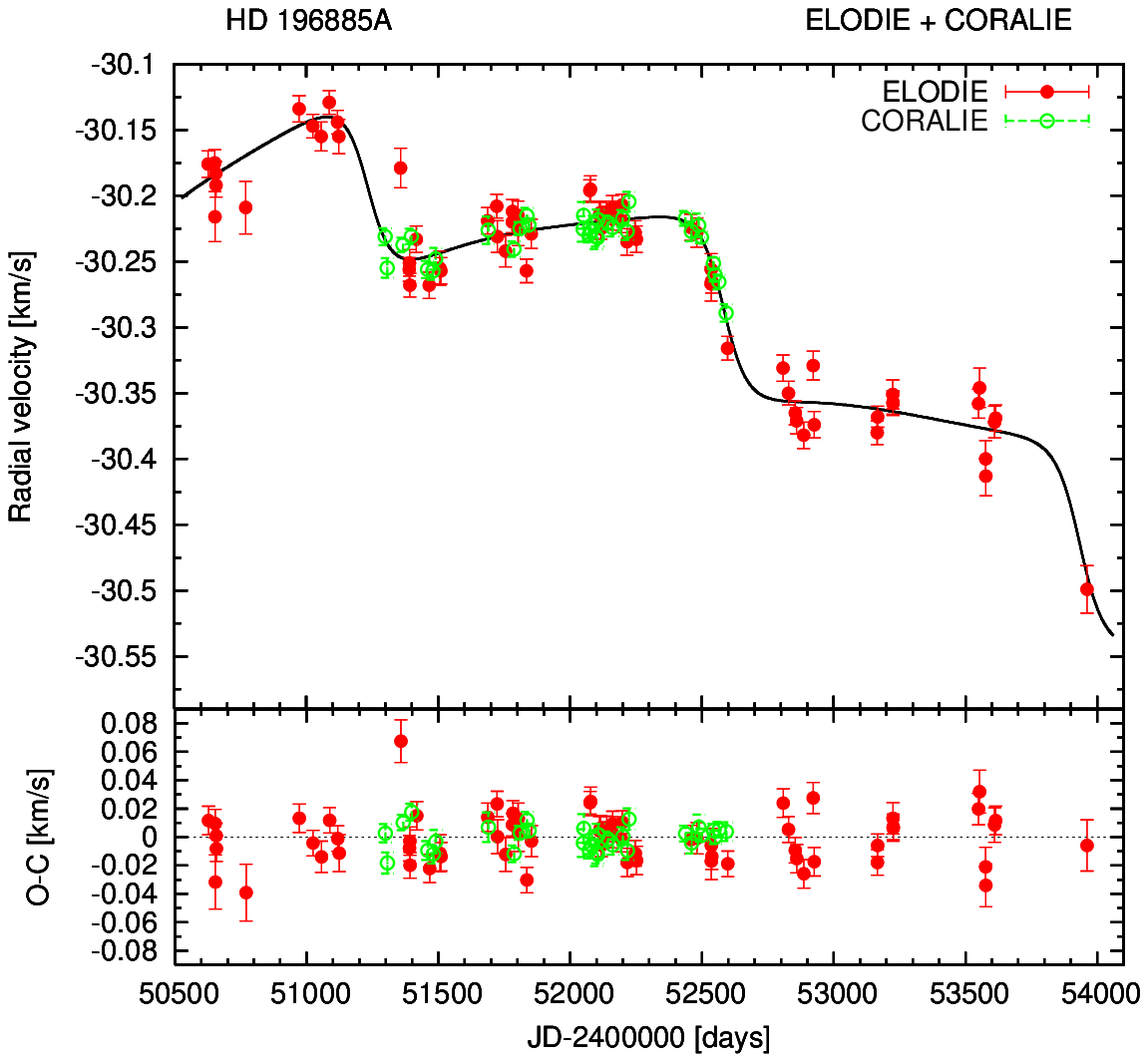}
  \caption{{\small ELODIE} and {\small CORALIE} radial velocities for {\small HD}\,196885\,A, 
    superimposed on a two-Keplerian orbital solution (Table~\ref{T2}).
    \label{F1} }
\end{figure}

The {\small ELODIE} observations of {\small HD}\,196885\,A started in
June 1997 and the last data acquired are from August 2006, since the {\small
ELODIE} program was closed shortly after that date.
The peculiar variations of the radial velocities (Fig.\ref{F1}) 
and also non-confirmed announcements from other research teams (see section
\ref{secondplanet}), prevented us from announcing this system earlier.
However, the recent detection of a visual small stellar companion close to the
main star \citep{Chauvin_etal_2006,Chauvin_etal_2007} confirmed our suspicion
of a long term drift of the radial velocities.
Superimposed on the drift we can also observe a regular variation of a few
years signaling the presence of a sub-stellar companion.

Before the {\small ELODIE} program, the star {\small HD}\,196885\,A had
been followed between June 1982 and August 1997 by the {\small CORAVEL}
spectrometers \citep{Baranne_etal_1979} mounted on the 1-m Swiss telescope at Haute-Provence Observatory
and on the 1.54-m Danish telescope at La Silla Observatory (ESO, Chile).
The precision of {\small CORAVEL} is $\sim$0.3~Km/s, not enough to detect
planetary objects, but very useful to help us to constrain the orbit
of the stellar companion.
From April 1999 to November 2002 a series of radial velocity measurements were
also taken using the {\small CORALIE} echelle spectrograph \citep{Queloz_etal_2000}
mounted on the 1.2-m Swiss telescope at La Silla.
This four year observational sequence is important to confirm the presence of the
planetary companion, since the precision of {\small CORALIE} is
slightly better than the precision of {\small ELODIE}.

\begin{figure}[t!]
   \centering
    \includegraphics*[height=8.5cm,angle=270]{\figpath 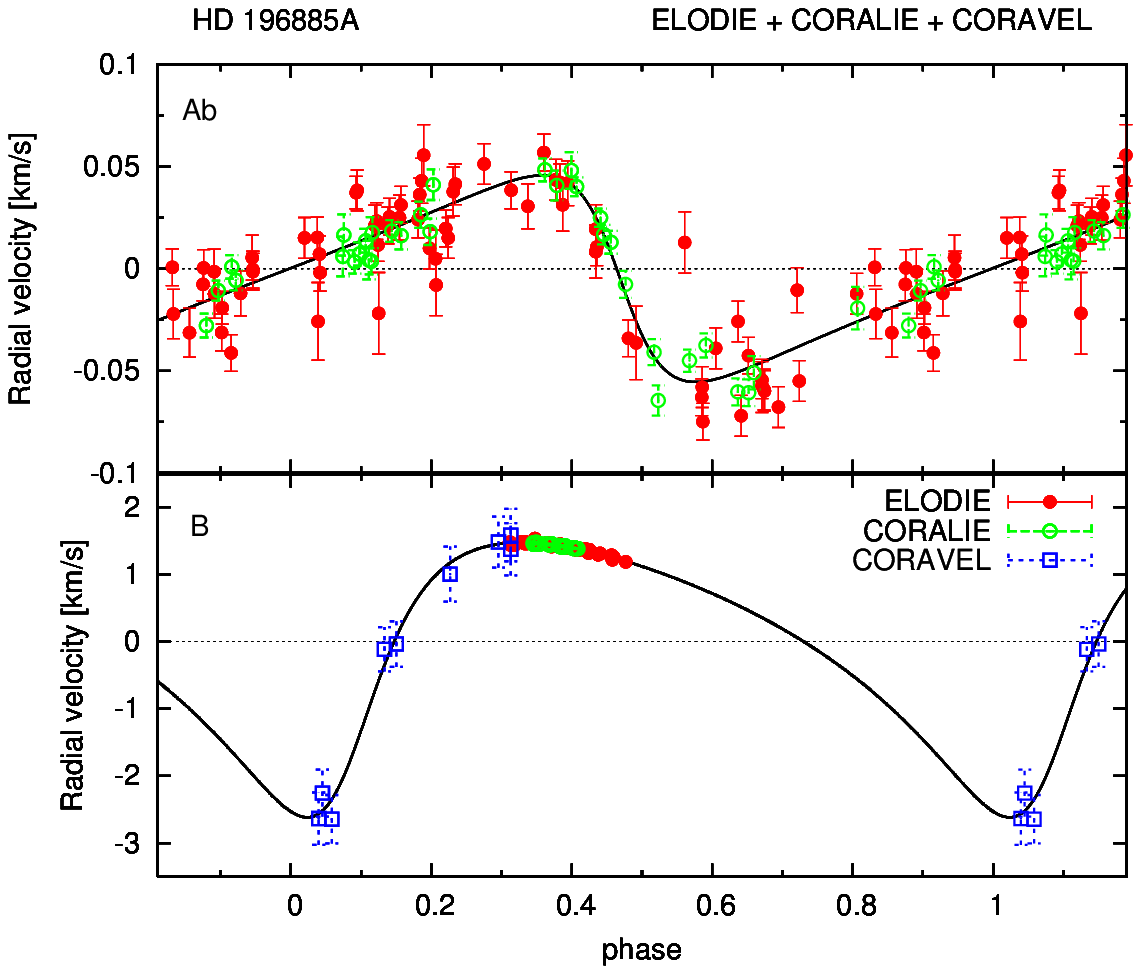}
  \caption{Phase-folded radial velocities measurements and best
  fit for the planetary companion of {\small HD}\,196885\,A (top) and the its
  stellar companion (bottom). For each body the contribution by the other
  companion has been subtracted from the observational data. The
  orbital period of the inner body is $P = 3.7$~yr, while for the outer body
  we have $P = 55$~yr (Table~\ref{T2}). However, the fact that we are unable to
  completely cover in phase the orbit of the stellar companion indicates
  that its orbital period may be considerably longer (Table~\ref{T3}). \label{F2} }
\end{figure}

With $\nptt$ radial-velocity measurements ($\npte$ from {\small ELODIE}, 9 from
{\small CORAVEL} and 33 from {\small CORALIE}), spanning $\sim$14~years of
observations, we are able to describe the orbit of the sub-stellar body in the
system, as well as slightly constrain the orbit of the stellar companion.
Using the iterative Levenberg-Marquardt method \citep{Press_etal_1992},
we first attempt to fit the complete set of radial velocities
with a single orbiting companion and a quadratic drift.  
This fit yields a planetary companion with $P=\perq$\,days, 
$e=\eccq$, a minimum mass of $\mssq\,M_\mathsf{Jup}$ and an adjustment of
$\sqrt{\chi^2}=\chiq$ and $rms=\rmsq\,\mathrm{m/s}$.
We then fit the radial velocities using a model with two Keplerian orbits
(Figs.\ref{F1} and \ref{F2}). 
It yields for the inner planet $P = \perd$~days, $e = 0.46$ and a minimum mass
of $3.0\,M_\mathsf{Jup}$, while for the outer companion we have
$P \simeq 20\,000$~days, $e = 0.41 $ and a minimum mass of
$\MssB\,M_\odot$ (Table~\ref{T2}). 
Despite all the uncertainties in the orbital parameters, the use of a Keplerian
orbit for the massive outer body in the system proved to be a good approach,
better than the quadratic drift, since the reduced $\sqrt{\chi^2}$ is now
$\chs$ and the velocity residuals drop to $rms$\,=\,$\wrms\,\mathrm{m/s}$
(Fig.\ref{F2}).
Of course one can always argue that increasing by four the number of free parameters in
the model will improve the fit, but at least we are able to place constraints for
the orbital parameters of the outer body.
For instance, we find that it must present an orbital period $ P >
40 $~yr, a semi-major axis $ a > 14 $\,AU and a minimum mass $ M \sin i > 0.28 \,
M_\odot$, since all other fits coherent with the data found larger values
for these parameters (Table~\ref{T3}).

\begin{table}[h!]
\caption{Orbital parameters for two bodies orbiting {\small
    HD}\,196885\,A, obtained with a two-Keplerian fit to
    observational data.}
\label{T2} 
\begin{center}
\begin{tabular}{l l c c} \hline \hline
\noalign{\smallskip}
{\bf Param.}  & {\bf [unit]} & {\bf {\small HD}\,196885\,Ab} & {\bf {\small HD}\,196885\,B} \\ \hline 
\noalign{\smallskip}
$V$          & [km/s]               & \multicolumn{2}{c}{$ \Vrad \pm 0.455 $}  \\  
$P$          & [year]              & $ \per \pm 0.03  $ & $ \perB \pm 19.48 $ \\ 
$e$          &                      & $ \ecc \pm 0.026 $ & $ \eccB \pm 0.038 $ \\ 
$\omega$     & [deg]                & $ \omg \pm 4.1   $ & $ \omgB \pm 23.4  $ \\ 
$K$          & [m/s]                & $ \Kap \pm 2.32  $ & $ \KapB \pm 270 $ \\  
$T$          & [JD-2400000]         & $ \Tpr \pm 18  $ & $ \TprB \pm 2638 $ \\  \hline
\noalign{\smallskip}
$a_1 \sin i$ & [$10^{-3}$ AU]       & $ \asn $           & $ \asnB $ \\
$f (M)$      & [$10^{-3}$ M$_\odot$]& $ \fm  $           & $ \fmB  $ \\
$M \sin i$ & [M$_\mathsf{Jup}$]   & $ \mss $           & $ \mssB $ \\
$M \sin i$ & [M$_\odot$]          & $ \Mss $           & $ \MssB $ \\
$a$          & [AU]                 & $ \sma $           & $ \smaB $ \\ \hline
\noalign{\smallskip}
$rms$        & [m/s]                & \multicolumn{2}{c}{\wrms}   \\
$\sqrt{\chi^2}$     &                      & \multicolumn{2}{c}{\chs}   \\ \hline
\noalign{\smallskip}
\end{tabular}
\end{center}
Errors are given by the standard deviation $ \sigma $.    
    The orbital period of the outer body is much longer than the
    data acquired so far ($\sim$14 yr) and thus we are unable to completely
    constrain its orbit. However, we noticed that it is better to fit a
    complete elliptical orbit to the present data, rather than use a quadratic
    drift, for which we obtained $\sqrt{\chi^2} = \chiq $ and $rms$ = \rmsq\,m/s.
    Alternative solutions for the outer body with longer orbital periods
    exist that match the data almost as well as the best fit (Table~\ref{T3}).
\end{table}

As expected, the orbital parameters of the outer body still present some
uncertainties around the best fitted value.
This is particularly true for the orbital period, meaning that this parameter may
assume rather different values.
The Levenberg-Marquardt method converges to a minimum $ \chi^2 $, but other
close local minima also may represent a good fit to our data.
For a particular choice of the orbital period, all the
other parameters can adjust in a way such that the $ \chi^2 $ value remains
close to the value from the best fit (Table~\ref{T2}).
Thus, any orbital solution with a $ \chi^2 $ and an $rms$ close to the values
from Table~\ref{T2} can be seen as a good model of the system
(eg. Table~\ref{T3}).

\begin{table}[h!]
\caption{Alternative orbital parameters for {\small HD}\,196885\,B.}
\label{T3} 
\begin{center}
\begin{tabular}{c | c c c c c | c c} \hline \hline
$ P $ & $ a $ & $ e $ & $ \omega $ & $ T $ & $ M \sin i $ & $rms$ & $ \sqrt{\chi^2} $ \\ 
$$ [yr] & [AU] &  & [deg] & [JD] & [$M_\odot$] & [m/s] & \\ \hline 
\noalign{\smallskip}
 40 & 13.7 & 0.50 & 204 & 45983 & 0.28 & 11.95 & 1.504 \\ 
 60 & 18.3 & 0.41 & 233 & 45916 & 0.36 & 11.87 & 1.494 \\ 
 80 & 22.4 & 0.42 & 250 & 46136 & 0.43 & 11.88 & 1.495 \\ 
100 & 26.6 & 0.45 & 258 & 45987 & 0.55 & 11.88 & 1.495 \\ 
120 & 30.4 & 0.49 & 265 & 46110 & 0.61 & 11.89 & 1.496 \\ \hline
\end{tabular}
\end{center}
With the present $\sim$14~years of observational
	data it is still impossible to completely constrain the outer body orbit.
	Observations by \citet{Chauvin_etal_2007} point to a small stellar companion
	with $0.5-0.6\,M_\odot$ at about 23\,AU. Fixing the orbital period at larger values, we
	are able to find orbital solutions compatible with those observations.
\end{table}

\citet{Chauvin_etal_2006,Chauvin_etal_2007} determined that {\small HD}\,196885
is a close visual binary.
They derive a possible mass for the companion of $0.5-0.6\,M_\odot$ at a
projected physical distance of $\sim$23\,AU and spectral type M1\,V. 
As they noticed, for those values of the mass and semi-major axis we should be
able to monitor its orbital motion using radial velocity measurements of the
main star.
In order to test if our drift can be interpreted as the observed stellar
companion, we fitted our data by fixing orbital periods compared to those
forecasted by \citet{Chauvin_etal_2007}.
As shown in Table~\ref{T3}, we are able to find multiple
possibilities that match their observations.
Notice also that the values of the mass listed in Table~\ref{T3} correspond to
minimum masses. Because of the factor $ \sin i $, the real masses can be larger.
We then conclude that the outer companion of the star {\small HD}\,196885\,A is
almost certainly the stellar mass object observed by
\citet{Chauvin_etal_2006,Chauvin_etal_2007}, denoted {\small HD}\,196885\,B.
The inner companion of {\small HD}\,196885\,A remains a planetary body,
hereafter called {\small HD}\,196885\,Ab, with orbital parameters
listed in Table~\ref{T2}.

Finally, we note that the {\small HD}\,196885 system may be a part of
a wider binary with the star {\small BD}+104251\,B located $\sim$191.9'' North of 
{\small HD}\,196885\,A \citep{Trilling_etal_2007}.
At a distance of 33\,pc measured by \citet{Perryman_etal_1997}, we estimate a
semi-major axis of 6330\,AU, that is, a drift contribution of $\sim$1\,$\mathrm{cm
\,s}^{-1}\mathrm{yr}^{-1}$ to the radial velocity of {\small HD}\,196885\,A
(assuming a circular orbit), and thus impossible to be detected with 
{\small ELODIE}. 
  

\section{A second planet around one year?}

\label{secondplanet}


In 2004, a $P=386$~days companion to the {\small HD}\,196885\,A
star was reported by the California \& Carnegie Planet Search Team on their
webpage\footnote{http://exoplanets.org/esp/hd196885/hd196885.shtml}, detected
through radial-velocity measurements obtained with the Lick survey.
This planet was also listed in the the exoplanets
Encyclopedia\footnote{http://exoplanet.eu/planet.php?p1=HD+196885\&p2=b}, by
\citet{Fischer_Valenti_2005}, \citet{Chauvin_etal_2007} and \citet{Marchi_2007}. 
Unfortunately no publication describing the system is known, as noticed by many other authors
\citep{Chauvin_etal_2006,Sousa_etal_2006,Bonavita_Desidera_2007,Desidera_Barbieri_2007}.
Moreover, the system is not reported in the 2006 exoplanet catalog by
\citet{Butler_etal_2006}, so we assume there was no confirmation of this planet
after the first announcement.

\begin{figure}[t!]
    \includegraphics*[height=8.5cm,angle=270]{\figpath 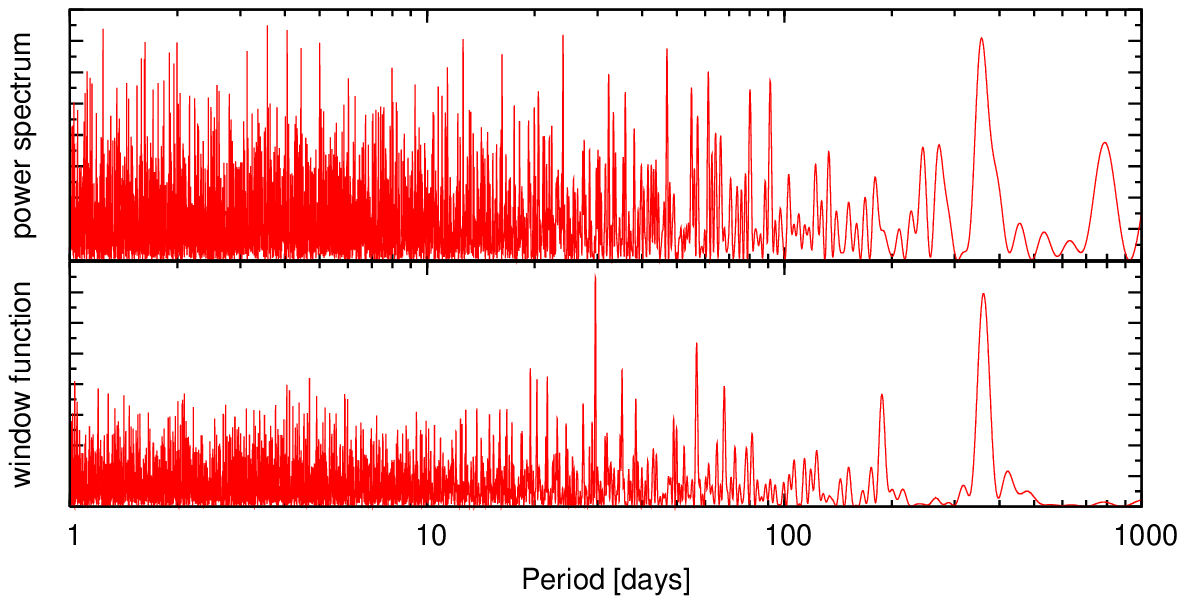} 
  \caption{Frequency analysis (top) and window function (bottom) for
  {\small ELODIE} residual radial velocities of {\small
  HD}\,196885\,A  when the contributions from the two
  bodies fitted in Table~\ref{T2} are subtracted. An important peak is detected at
  $P = 368$\,days, which could be interpreted as a third body in the system.
  However, looking at the window function, we see that the same peak is present,
  indicating that this may be an artifact. %
  \label{F3}}   
\end{figure}

\begin{figure}[t!]
    \includegraphics*[height=8.5cm,angle=270]{\figpath 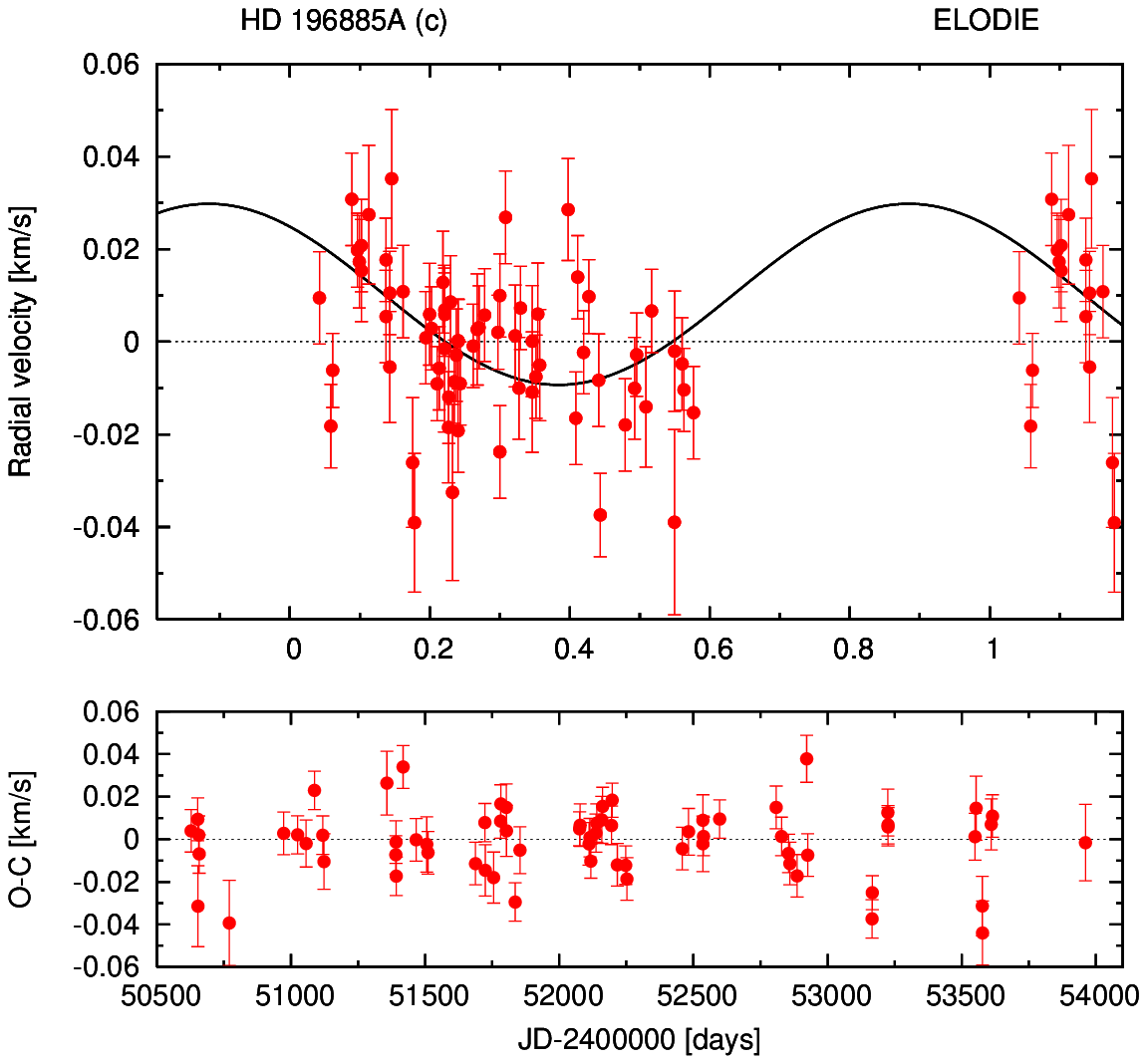} 
  \caption{Phase-folded {\small ELODIE} residual radial velocities for 
    {\small HD}\,196885\,A when the contributions from the two bodies fitted
    in Table~\ref{T2} are subtracted. Data are superimposed on a Keplerian solution
    with $ e = 0 $ and $P = 368$\,days. The respective residuals as a function of
    Julian Date are displayed in the lower panel. We see that {\small ELODIE} data only cover half of the
    orbital phase, preventing us from completely constraining it. Moreover, the inclusion of this
    third body does not improve the fit significantly. 
    We then conclude that the presence of a third body at $ P = 368 $\,days is
    an artifact. 
    \label{F4}}   
\end{figure}

Curiously, performing a frequency analysis of the {\small ELODIE} radial
velocity residuals (Fig.\ref{F1}), we find an important peak signature at about $
P = 368 $~days (Fig.\ref{F3}), that could be interpreted as a second planet
around {\small HD}\,196885\,A.
However, this peak is not present when we analyze {\small CORALIE} data, casting
some doubts on its origin. 
Computing the {\small ELODIE} window function, we also note an important peak at exactly $ 368
$~days (Fig.\ref{F3}), clearly suggesting that the peak shown in the frequency
analysis is an aliasing of the observational data.
Performing a Keplerian fit to the {\small ELODIE} residuals, we find that
the best fit corresponds to solutions with very high eccentricities ($e
\sim 0.8 $), that are not dynamical stable.
Moreover, when we plot a phase-fold of the residual radial velocities
(Fig.\ref{F4}), our observations only cover half of the orbit,
preventing us from fully constraining it.
Finally, we performed a Keplerian fit of the three bodies also including data from
{\small CORALIE} and {\small CORAVEL}.
We found $ \sqrt{\chi^2} = 1.460 $ and $rms$\,=\,$11.30\,\mathrm{m/s}$,
which does not represent a substantial improvement with respect to
the system with only two companions (Table~\ref{T2}).
The presence of a companion around $ 368 $~days can thus be discarded.

Besides the artifactual companion at $ 368 $~days, we may ask if there are
other companions at different orbital periods.
For that purpose we used a genetic algorithm, since we cannot
clearly isolate any other peak in the frequency analysis of Figure~\ref{F3}.
The inclusion of a third companion in the system allows us to reduce the $
\sqrt{\chi^2} $ to 1.25 and the $rms$\,=\,$10.5\,\mathrm{m/s}$.
This represents only a slightly better adjustment of the model to the data that
can be justified as a natural consequence of increasing the number of free
parameters. 
Moreover, identical adjustments can be obtained with many orbital periods, as
different as $ 1.3$~days and $ 10 $~days, 
frequently with very high eccentricity values.
Therefore, no other companion can be conclusively detected in the
residuals from the orbital solution listed in Table~\ref{T2}.


\section{Discussion and conclusion}

In this paper we report the presence of a planet orbiting the
{\small HD}\,196885\,A star, with an orbital period of \perd~days. 
We also detect the presence of the stellar companion {\small HD}\,196885\,B, that
was recently observed by \citet{Chauvin_etal_2006,Chauvin_etal_2007}.
Its orbit is not completely determined, but we are able to provide some
constraints: $ P > 40 $~yr, $ a > 14 $\,AU and $ M \sin i > 0.28 \, M_\odot$.
It is possible to
find a large set of orbits with much longer orbital periods that fit the
observational data as well as the minimum mass orbits (Table~\ref{T3}). 
In particular, we find a solution at $ P = 120 $~yr with $ M \sin i = 
0.6\,M_\odot $, the mass estimated by \citet{Chauvin_etal_2007}. 

The {\small HD}\,196885\,A system was previously described as a star orbited by a
planet at 386~days by the California \& Carnegie Planet Search Team, detected with
data acquired with the Lick survey. 
Although the system is reported on their webpage and many other
places, no publication describing the system is yet known.
The data acquired with the {\small ELODIE} survey also present some signal around $
368 $~days, but a more detailed analysis shows that this signal corresponds to
an aliasing of the observational data, probably due to the similarities with the
Earth orbital period. 
We thus do not confirm the existence of the planet obtained with measurements form
the Lick survey.

The {\small ELODIE} radial velocity residuals after subtracting the signal from
the two companions still presents an $O-C$ (Fig.\ref{F1}) slightly above the
precision of the instrument.
This suggests that the system may still hide additional planetary companions.
However, we were unable to find them, even when using a genetic algorithm.
The excess in the residuals can also be a contamination 
from the  spectrum of the stellar companion {\small HD}\,196885\,B.
We have searched for its presence
in our {\small ELODIE} data using the multi-order TODCOR, a two-dimensional
cross-correlation algorithm \citep{Zucker_etal_2004},
but did not find anything convincing.
We have also searched for correlations between line bisectors and radial
velocity or residuals. 
No significant correlation could be detected. 
These negative results do not allow us to formally discard the blend scenario,
but they render it an unlikely possibility.

The planet {\small HD}\,196885\,Ab is the fourth to be discovered in a close
binary with a separation smaller than 25\,AU.
The other cases already known are {\small Gliese}\,86
\citep{Queloz_etal_2000,Lagrange_etal_2006}, $\gamma$ Cep
\citep{Hatzes_etal_2003,Neuhaeuser_etal_2007} and {\small 
HD}\,41004\,A \citep{Zucker_etal_2004}.
These systems are ideal for carrying out combined astrometric and
radial-velocity observations to
constrain the binary dynamic properties and the possible impact
of a close binary companion on planet formation and evolution.

\vskip0.5cm

\begin{acknowledgements}
We acknowledge support from the Swiss
National Research Found (FNRS), the Geneva University and
French CNRS. A.C. and N.S. benefited from Portuguese FCT grant
POCI/CTE-AST/56453/2004 and N.S. was also supported from the EC's FP6 and by FCT
(with POCI2010 and FEDER funds), within the HELAS international
collaboration. A.E. acknowledges support from the Swiss National
Science Foundation through a fellowship for a prospective researcher.
We are grateful to the Observatoire de Haute-Provence 
for the generous time allocation.
\end{acknowledgements}

{\bf Note:} After the submission of the present paper the web sites were updated with
the orbital solution described here. Moreover, the Lick
Planet Search Team also agree with this new solution for {\small HD}\,196885
(Debra Fisher, private communication).

\bibliographystyle{aa}
\bibliography{correia}

\begin{thebibliography}{28}
\expandafter\ifx\csname natexlab\endcsname\relax\def\natexlab#1{#1}\fi

\bibitem[{{Baranne} {et~al.}(1979){Baranne}, {Mayor}, \&
  {Poncet}}]{Baranne_etal_1979}
{Baranne}, A., {Mayor}, M., \& {Poncet}, J.~L. 1979, Vistas in Astronomy, 23,
  279

\bibitem[{{Baranne} {et~al.}(1996){Baranne}, {Queloz}, {Mayor}, {Adrianzyk},
  {Knispel}, {Kohler}, {Lacroix}, {Meunier}, {Rimbaud}, \&
  {Vin}}]{Baranne_etal_1996}
{Baranne}, A., {Queloz}, D., {Mayor}, M., {et~al.} 1996, \aaps, 119, 373

\bibitem[{{Bonavita} \& {Desidera}(2007)}]{Bonavita_Desidera_2007}
{Bonavita}, M. \& {Desidera}, S. 2007, \aap, 468, 721

\bibitem[{{Bouchy} \& {The Sophie Team}(2006)}]{Bouchy_etal_2006}
{Bouchy}, F. \& {The Sophie Team}. 2006, in Tenth Anniversary of 51 Peg-b:
  Status of and prospects for hot Jupiter studies, ed. L.~{Arnold},
  F.~{Bouchy}, \& C.~{Moutou}, 319--325

\bibitem[{{Butler} {et~al.}(2006){Butler}, {Wright}, {Marcy}, {Fischer},
  {Vogt}, {Tinney}, {Jones}, {Carter}, {Johnson}, {McCarthy}, \&
  {Penny}}]{Butler_etal_2006}
{Butler}, R.~P., {Wright}, J.~T., {Marcy}, G.~W., {et~al.} 2006, \apj, 646, 505

\bibitem[{{Chauvin} {et~al.}(2006){Chauvin}, {Lagrange}, {Udry}, {Fusco},
  {Galland}, {Naef}, {Beuzit}, \& {Mayor}}]{Chauvin_etal_2006}
{Chauvin}, G., {Lagrange}, A.-M., {Udry}, S., {et~al.} 2006, \aap, 456, 1165

\bibitem[{{Chauvin} {et~al.}(2007){Chauvin}, {Lagrange}, {Udry}, \&
  {Mayor}}]{Chauvin_etal_2007}
{Chauvin}, G., {Lagrange}, A.~M., {Udry}, S., \& {Mayor}, M. 2007, \aap, 475,
  723

\bibitem[{{Desidera} \& {Barbieri}(2007)}]{Desidera_Barbieri_2007}
{Desidera}, S. \& {Barbieri}, M. 2007, \aap, 462, 345

\bibitem[{{ESA}(1997)}]{ESA_1997}
{ESA}. 1997, The Hipparcos and Tycho Catalogues, 1239

\bibitem[{{Fischer} \& {Valenti}(2005)}]{Fischer_Valenti_2005}
{Fischer}, D.~A. \& {Valenti}, J. 2005, \apj, 622, 1102

\bibitem[{{Flower}(1996)}]{Flower_1996}
{Flower}, P.~J. 1996, \apj, 469, 355

\bibitem[{{Hatzes} {et~al.}(2003){Hatzes}, {Cochran}, {Endl}, {McArthur},
  {Paulson}, {Walker}, {Campbell}, \& {Yang}}]{Hatzes_etal_2003}
{Hatzes}, A.~P., {Cochran}, W.~D., {Endl}, M., {et~al.} 2003, \apj, 599, 1383

\bibitem[{{Lagrange} {et~al.}(2006){Lagrange}, {Beust}, {Udry}, {Chauvin}, \&
  {Mayor}}]{Lagrange_etal_2006}
{Lagrange}, A.-M., {Beust}, H., {Udry}, S., {Chauvin}, G., \& {Mayor}, M. 2006,
  \aap, 459, 955

\bibitem[{{Marchi}(2007)}]{Marchi_2007}
{Marchi}, S. 2007, \apj, 666, 475

\bibitem[{{Mayor} \& {Queloz}(1995)}]{Mayor_Queloz_1995}
{Mayor}, M. \& {Queloz}, D. 1995, \nat, 378, 355

\bibitem[{{Neuh{\"a}user} {et~al.}(2007){Neuh{\"a}user}, {Mugrauer},
  {Fukagawa}, {Torres}, \& {Schmidt}}]{Neuhaeuser_etal_2007}
{Neuh{\"a}user}, R., {Mugrauer}, M., {Fukagawa}, M., {Torres}, G., \&
  {Schmidt}, T. 2007, \aap, 462, 777

\bibitem[{{Pace} \& {Pasquini}(2004)}]{Pace_Pasquini_2004}
{Pace}, G. \& {Pasquini}, L. 2004, \aap, 426, 1021

\bibitem[{{Perrier} {et~al.}(2003){Perrier}, {Sivan}, {Naef}, {Beuzit},
  {Mayor}, {Queloz}, \& {Udry}}]{Perrier_etal_2003}
{Perrier}, C., {Sivan}, J.-P., {Naef}, D., {et~al.} 2003, \aap, 410, 1039

\bibitem[{{Perryman} {et~al.}(1997){Perryman}, {Lindegren}, {Kovalevsky},
  {Hoeg}, {Bastian}, {Bernacca}, {Cr{\'e}z{\'e}}, {Donati}, {Grenon}, {van
  Leeuwen}, {van der Marel}, {Mignard}, {Murray}, {Le Poole}, {Schrijver},
  {Turon}, {Arenou}, {Froeschl{\'e}}, \& {Petersen}}]{Perryman_etal_1997}
{Perryman}, M.~A.~C., {Lindegren}, L., {Kovalevsky}, J., {et~al.} 1997, \aap,
  323, L49

\bibitem[{{Pont} \& {Eyer}(2004)}]{Pont_Eyer_2004}
{Pont}, F. \& {Eyer}, L. 2004, \mnras, 351, 487

\bibitem[{{Press} {et~al.}(1992){Press}, {Teukolsky}, {Vetterling}, \&
  {Flannery}}]{Press_etal_1992}
{Press}, W.~H., {Teukolsky}, S.~A., {Vetterling}, W.~T., \& {Flannery}, B.~P.
  1992, {Numerical recipes in FORTRAN. The art of scientific computing}
  (Cambridge: University Press, 2nd ed.)

\bibitem[{{Queloz} {et~al.}(2000){Queloz}, {Mayor}, {Weber}, {Bl{\'e}cha},
  {Burnet}, {Confino}, {Naef}, {Pepe}, {Santos}, \& {Udry}}]{Queloz_etal_2000}
{Queloz}, D., {Mayor}, M., {Weber}, L., {et~al.} 2000, \aap, 354, 99

\bibitem[{{Schaerer} {et~al.}(1993){Schaerer}, {Meynet}, {Maeder}, \&
  {Schaller}}]{Schaerer_etal_1993}
{Schaerer}, D., {Meynet}, G., {Maeder}, A., \& {Schaller}, G. 1993, \aaps, 98,
  523

\bibitem[{{Schaller} {et~al.}(1992){Schaller}, {Schaerer}, {Meynet}, \&
  {Maeder}}]{Schaller_etal_1992}
{Schaller}, G., {Schaerer}, D., {Meynet}, G., \& {Maeder}, A. 1992, \aaps, 96,
  269

\bibitem[{{Sousa} {et~al.}(2006){Sousa}, {Santos}, {Israelian}, {Mayor}, \&
  {Monteiro}}]{Sousa_etal_2006}
{Sousa}, S.~G., {Santos}, N.~C., {Israelian}, G., {Mayor}, M., \& {Monteiro},
  M.~J.~P.~F.~G. 2006, \aap, 458, 873

\bibitem[{{Trilling} {et~al.}(2007){Trilling}, {Stansberry}, {Stapelfeldt},
  {Rieke}, {Su}, {Gray}, {Corbally}, {Bryden}, {Chen}, {Boden}, \&
  {Beichman}}]{Trilling_etal_2007}
{Trilling}, D.~E., {Stansberry}, J.~A., {Stapelfeldt}, K.~R., {et~al.} 2007,
  \apj, 658, 1289

\bibitem[{{Wright} {et~al.}(2004){Wright}, {Marcy}, {Butler}, \&
  {Vogt}}]{Wright_etal_2004}
{Wright}, J.~T., {Marcy}, G.~W., {Butler}, R.~P., \& {Vogt}, S.~S. 2004, \apjs,
  152, 261

\bibitem[{{Zucker} {et~al.}(2004){Zucker}, {Mazeh}, {Santos}, {Udry}, \&
  {Mayor}}]{Zucker_etal_2004}
{Zucker}, S., {Mazeh}, T., {Santos}, N.~C., {Udry}, S., \& {Mayor}, M. 2004,
  \aap, 426, 695

\end{thebibliography}

\end{document}